\documentstyle[12pt]{article}

\begin{document}

\begin{center}

\null
\vskip-1truecm
\rightline{IC/93/63}
\vskip1truecm

{International Atomic Energy Agency\\
and\\
United Nations Educational Scientific and Cultural Organization\\
\medskip
INTERNATIONAL CENTRE FOR THEORETICAL PHYSICS\\}
\vskip2truecm

{\bf TOPOLOGICALLY STABLE ELECTROWEAK
FLUX TUBES}

\vskip2truecm
{Gia Dvali \footnote{e-mail:
Dvali@ITSICTP.Bitnet}\quad  and \quad Goran
Senjanovi\'c \footnote{e-mail: Goran@ITSICTP.Bitnet}\\
International Centre for Theoretical Physics, Trieste, Italy.\\}

\vskip1.5truecm
\begin{abstract}
\baselineskip=20pt

We show that for a large range of parameters in a
$SU(2)_L\times U(1)$  electroweak theory with two
Higgs doublets there may exist classically stable flux
tubes of Z boson magnetic field.  In a limit of an extra
global $\tilde U(1)$     symmetry, these flux-tubes become
topologically stable.  These results are automatically
valid even if $\tilde U(1)$     is gauged.
\end{abstract}

\vskip2truecm

{MIRAMARE--TRIESTE\\
\bigskip
April 1993\\}

\end{center}

\newpage

\baselineskip = 25pt

\noindent{\bf A.\quad Introduction}
\bigskip

More than twenty years have passed since
 Nielsen and Olesen discovered that Abrikosov type
vertices may appear as classical solutions in
spontaneously broken gauge theories [1].   It is expected
that these objects, so called strings, play an important
role in both particle physics and cosmology [2].  It is of
crucial importance to know if strings can exist in
the   $SU(2)_L\times U(1)$              based electroweak
theory.  Recently, in an inspiring paper Vachaspati showed
that classically stable string-like structures can exist
even in the standard model [3] (but unfortunately only for
unrealistic values of  $\sin^2\theta_W$         and Higgs
mass [4]).  These are the usual Nielsen-Olesen flux tubes
embedded in  $SU(2)_L\times U(1)$         group and
therefore no longer topologically stable.

	In this paper we investigate under which
conditions Z boson flux tubes could be actually
topologically stable.  Much to our surprise, it turns out
that the price needed to achieve this is just the
additional global   $\tilde U(1)$   symmetry in the Higgs
sector. The minimal structure that can lead to
extra $\tilde U(1)$       is a two Higgs
doublet             model, with these fields carrying
different  $\tilde U(1)$   charges.  As far as our analysis is concerned,
this symmetry may or may not be anomalous.
Depending on the choice of the fermionic charge
assignment, this symmetry can be identified with a
Peccei-Quinn [5] or B-L symmetry.  We return to the issue
of realistic models in part C; now without further ado we
wish to demonstrate the nature of this phenomenon.
\bigskip

\noindent{\bf B.\quad Global Symmetry and Topologically Stable
Z-Flux Tubes}
\bigskip

	Imagine a   $SU(2)_L\times U(1)$          electroweak
gauge theory with two doublets $\Phi_1$   and $\Phi_2$
and a potential
$$V = \sum^2_{i=1} {\lambda_i\over 4}
(\Phi^+_i\Phi_i-m^2_i)^2+{\lambda\over
2}(\Phi^+_1\Phi_1)(\Phi^+_2\Phi_2)+{\lambda'\over
2}(\Phi^+_1\Phi_2)(\Phi^+_2\Phi_1)\eqno(1)$$

 Obviously, the
above potential is the most general one invariant
under  $SU(2)_L\times U(1)$             gauge symmetry and
an additional global $\tilde U(1)$     symmetry.  The extra
symmetry simply amounts to the freedom of independent phase
transformations of the two doublets.

	It is clear that in the range of parameters
(which includes   $\lambda'<0$), both  $\Phi_i\ \  (i=1,2)$
develop vacuum expectation values in one and the same
direction in the group space,
$$\langle \Phi_i\rangle =\left(\matrix{0\cr
v_i\cr}\right) e^{i\theta_i},\quad
0\leq \theta_i\leq 2\pi\eqno(2)$$
which preserves  $U(1)_{em}$      much
as in the standard model:
$$G\equiv SU(2)_L\times U(1)\times \tilde
U(1)\mathop{\longrightarrow}\limits_{v_i\not= 0}
U(1)_{em}\equiv H\eqno(3)$$

 Notice that both $v_i$      have to
be nonvanishing in order for $\tilde U(1)$      global
symmetry to be broken.  Obviously, the vacuum manifold is
not simply connected ($\pi_1(G/H)\not= 1$) and therefore,
as is well-known, there must be a topologically stable
string solution in this theory.  The origin of the string
is clear, since the potential includes no couplings
explicitly dependent on the phase difference
$\theta\equiv \theta_1-\theta_2$.  Therefore,
$\theta_1$      and $\theta_2$      are locally
uncorrelated so that     $\theta$  can wind by
$2\pi n$ ($n$     is an integer) around some closed path,
ensuring the existence of the string.  The stability of
the string is guaranteed by the topologically invariant
condition
$${1\over 2\pi} \oint dx^\mu \partial_\mu\theta =
n\eqno(4)$$where integration is carried over the path
enclosing the string at infinity.  Naively, one imagines
this string to be global since it results from the
spontaneous breaking of a global   $\tilde U(1)$
factor.

	In other words, this string is not expected
to carry any magnetic flux associated
with   $SU(2)_L\times U(1)$                 group.
However, this premise is false as we now demonstrate.

	The crucial point is that besides the usual
topologically invariant boundary condition (4), one must
also demand the single valuedness of the vev's
$v_i$, i.e. the phases   $\theta_i$
must return to their original values while encircling the
string.  More precisely, this implies topological
constraints
$${1\over 2\pi} \oint dx^\mu \partial_\mu
\theta_i = n_i\eqno(5)$$
where $n_i$     are integers, and so
from (4) one has $n=n_1-n_2$.  In what follows, we
consider a single string solution, i.e. $n=1$ case.

	Using the equations of motion
 for the Z boson field, one easily obtains the behaviour
far away from the core of the string
$$Z_\mu ={\cos\theta_W\over g}
{v^2_1\partial_\mu \theta_1+v^2_2\partial_\mu
\theta_2\over v^2_1+v^2_2}\eqno(6)$$

 Integrating (6) over the same loop as before,
we obtain
$$\oint Z_\mu dx^\mu ={\cos\theta_W\over g}
{v^2_1n_1+v^2_2n_2\over v^2_1+v^2_2}\eqno(7)$$

 Once again,
since $n_i$       are integers, the right--hand side of (7)
can never vanish, which implies the nonvanishing Z boson
flux, trapped in the string.
	This is the key observation of our paper.

	Note that the Z flux in (7) differs from the
original one of Nielsen and Olesen, since as long
as $v^2_1\not= v^2_2$          (as expected in realistic
situations) it will not be an integer.  However, it is
topologically stable, due to the topological nature of the
scalar configuration behind its existence.

	We wish to note here that our expression (6) and
the non-integral nature of the Z flux is reminiscent of
the situation encountered by Hill et al. in their study of
so called frustrated strings [6].  Their philosophy was,
however, orthogonal to ours since they attributed the
existence of strings to the spontaneous breaking of a
local gauge symmetry and made no use of a global
symmetry (or at least show no
awareness of it).  Even more important, in their work there is no mention of
possible relevance of their results for the physically interesting
case of electroweak Z flux tubes.
\bigskip

\noindent{\bf C.\quad Realistic Models}
\bigskip

	This is all very nice, but unfortunately the resulting
 Goldstone boson due to   $\tilde U(1)$
breaking is necessarily coupled to fermions, independently
of whether both or just one Higgs doublet is coupled to
fermions.  Therefore, stellar objects would copiously
produce such particles and radiate their energy away
unless their couplings to light fermions were strongly
suppressed, typically being less than   $10^{-10}$
[7].  As is well known, the way out is to introduce
a   $SU(2)_L\times U(1)$             singlet field S,
by attributing              breaking  of $\tilde U(1)$ to
its large expectation value
$v_S\equiv \vert\langle
S\rangle\vert\geq 10^9$ GeV.
A typical example is a celebrated invisible axion scenario
[8] with    $\tilde U(1)$            being a Peccei-Quinn
symmetry.  The  way to define   $\tilde
U(1)$           charge of S is through the explicit phase
dependent coupling in the potential
$$\Delta V({\rm phase}) =\mu (\Phi^+_1\Phi_2 S + {\rm
h.c.}) =\mu v_1v_2\cos(\theta_S-\theta)\eqno(8)$$

 Clearly,
as soon as S picks up a non zero vev , it breaks
$\tilde U(1)$     and a global string is formed through
the winding of $\theta_S$    by   $2\pi$   (again we
restrict ourselves to  $n=1$      case).  Next when
$\Phi_i$        develop vev's and
break   $SU(2)_L\times U(1)$
 their phases
are locally correlated to  $\theta_S$             through
the coupling in (8), i.e. $\theta=\theta_S$,
or  $n_1-n_2=1$.  As a consequence, our previous
discussion becomes automatically valid.  Notice an
important point that Z flux in (6) is completely
independent of the global symmetry breaking scale.

	Now that we have seen that our solution
is for real, let us discuss some natural candidates
for  $\tilde U(1)$.

\noindent (a)	Peccei-Quinn (axion) case

	In this case the doublets  $\Phi_i$        couple
separately to up and down quarks.  Resulting axion strings
have been studied at length and as is well known, due to
instanton induced explicit breaking of
$U(1)_{PQ}$             these strings become boundaries of
domain walls below the scale of QCD phase transition [9].
The structures decay rapidly before dominating the energy
content of the universe (unless there are truly stable
domain walls) [9].

\noindent
(b)	Majoron case

	Another natural candidate for    $\tilde U(1)$       that
comes to mind is B-L symmetry which is automatically
preserved in the standard model.  The point is that B-L
symmetry is free from anomalies which destabilise the Z
flux tubes.

	Unfortunately, the simplest and most popular
Majoron scheme leaves no room for the considered
structures since it is based on a single Higgs doublet
picture [10].  One simply adds a singlet S and a
right-handed neutrino   $\nu_R$, and defines the B-L
of S through the coupling
$$L_\gamma(s)=h_R S\nu^T_R C \nu_R + {\rm
h.c.}\eqno(9)$$
where the B-L property of the right handed neutrino
is defined through
$$L_\gamma(\nu) =h_\nu(\bar\nu \bar e)_L \Phi^*\nu_R +
{\rm h.c.}\eqno(10)$$

Clearly, one Higgs doublet suffices
and it carries no B-L quantum number.  In turn the B-L
strings are purely global and devoid of any Z flux.  We
show how this can be easily modified
in a two doublet case.

\noindent
(c)	Anomaly Free Two Higgs Doublet Model

	There is an unique version of two doublet model which has
no global anomaly, one with one of the doublets decoupled
from the quarks.  In addition, this automatically ensures
natural flavour conservation in neutral currents.

	The global symmetry in this case can be identified in the
fermion sector as the global hyperchange rotation which
leaves the decoupled Higgs doublet, say  $\Phi_2$,
invariant.  Clearly, this symmetry is anomaly free.
However, since it is chiral and the corresponding
Goldstone boson has diagonal couplings to light fermions,
once again one is led to introduce a singlet coupled as
in (8).  As before, consistency, of the model
requires   $v_S\geq 10^9$ GeV.                The
distinguishing feature with respect to the axion model is,
as we emphasised before, the topological stability of the
Z flux tubes due to the absence of anomalies.  The strings,
therefore, will never become boundaries of domain walls.

	The nice feature of this scenario is that the global
 freedom of the model can take the meaning of B-L.  All
one has to do is to couple S to   $\nu_R$      in (9), and
the        charge becomes a liner superposition of
hypercharge Y and B-L
$$\tilde Q =2Y -5(B-L)\eqno(11)$$

	The unusual feature of
this picture is that Majoron is now coupled at the tree
level to both the electron and the light quarks, but its
couplings are suppressed by $v^{-1}_S$.
\newpage

\noindent {\bf D.	\quad   $\tilde U(1)$   Breaking and Z Flux Tubes}
\bigskip

	The essential ingredient in our construction
of Z flux tubes was the existence of a global symmetry.
What happens when this symmetry is explicitly broken?
Here we wish to show that Z flux can remain classically
stable in a certain range of parameters of the theory even
if $\tilde U(1)$      global symmetry is explicitly
violated and there is no ultra light pseudoscalar boson in
the spectrum.  In this case Z flux will flow along the
boundary of the domain wall that terminates in its tube.

	To see what happens, let us switch
 on an explicit  $\tilde U(1)$    breaking in the two
Higgs doublet model
$$\Delta V ={\gamma\over 4}[(\Phi^+_1\Phi_2)^2+{\rm
h.c.}]={\gamma\over 2} v^2_1 v^2_2\cos
2\theta\eqno(12)$$
For simplicity we restrict ourselves to the
above term only which automatically preserves
a discrete  $Z_2$ symmetry:
$\Phi_1\to -\Phi_1$,
$\Phi_2\to \Phi_2$   (or vice
versa), needed to ensure natural flavour conservation in
the neutral currents.  Now, as soon as  $\gamma\not=
0$,the phases $\theta_1$    and $\theta_2$    become
locally correlated.  E.g., for  $\gamma<0$   (12)
is minimized for $\theta= 0$,
however   $\theta$        cannot vanish everywhere, since
it has to wind up around the string.  In other words, while
enclosing the string, one is forced to pass through a
region   $\theta\not= 0$           indicating the existence
of a domain wall attached to a string.  This string-wall
system will remain classically stable as long as there
exists a potential barrier that prevents the unwinding of
this configuration.

	Choosing say  $v_1\gg v_2$, tells us that Z flux
tube will be classically stable, as long as
nonvanishing    $v_2$        is energetically favoured
everywhere in space (including the vicinity of the domain
wall), or in other words, as long as the
(mass)$^2$            of  $\vert\Phi_2\vert$            is
not always negative.  The contribution to this term comes
from the potential and the gradient energy (the latter is
significant only inside the wall).  The effective mass
term has the form
$$m^2_{eff}=-\lambda_2m^2_2+(\lambda+\lambda'+\gamma\cos
2\theta)v^2_1+\bigg\vert{\partial\theta\over \partial
x}\bigg\vert^2\eqno(13)$$
where x is the coordinate transverse
to the wall.  Obviously, the string-wall system can be
unstable only if $m^2_{eff}$          becomes positive at
some point inside the wall pushing
$\vert\Phi_2\vert$    over the top of the
Mexican hat potential.

	Note that the  $\tilde U(1)$
                    violating expression (12) gives a mass
to a would have been Goldstone boson
$$m^2_a\simeq \gamma v^2_1\eqno(14)$$
where the subscript
refers to an "axionic" nature of the particle.

	Furthermore,
since $\theta(x)$       changes by $\Delta\theta =
\pi$        through the wall whose thickness
is $\delta\sim m^{-1}_a$             one
gets
$\vert{\partial\theta\over\partial
x}\vert\simeq \pi m_a$.  In terms of
physical fields, the condition $m^2_{eff}<0$ implies that
the "axion" field is lighter than the radial mode of the
string.

	Of course, although classically stable
this system can decay through quantum mechanical tunneling
via the hole formation in the wall sheet.  However, this
decay rate will be exponentially suppressed by the ratio
of the mass of the radial mode to the mass of the
"axion".  Clearly, we end up predicting that the "axion"
mass should lie below   $M_W$      in order for the
considered structure to be stable.

	In addition, if we switch on
the other possible phase dependent couplings in the
potential of the type $\Phi^+_1\Phi_2$, $Z_2$
       symmetry gets explicitly broken and the
two  $\Delta \theta=\pi$             domain walls attached
to the string will collapse into the single one
with      $\Delta\theta
=2\pi$.

	For simplicity, you can  assume the coefficients to be
real.  This offers an even more interesting possibility of
spontaneous CP violation.  Therefore, previous analysis
shows an intriguing possibility of CP domain walls
terminating into Z flux tubes.
\bigskip

\noindent {\bf E.\quad	Conclusion and Outlook}
\bigskip

	As we mentioned in the Introduction,
Vachaspati has shown that contrary to the conventional
wisdom, there may exist classically stable string-like
structures in the standard electroweak model, but,
unfortunately not in the realistic range of parameters.
The main problem in his construction lies in the fact that
the usual Nielsen-Olesen flux tubes embedded
in  $SU(2)_L\times U(1)$          are not topologically
stable.  This leads one naturally to pursue the situation
when there is a global $\tilde U(1)$       symmetry in the
theory, for then the existence of (global) strings is
ensured by topological considerations.  And if the
relevant scalar fields interact with the Z boson, this
will lead to the trapping of the Z flux on these
string-like structures.  This is the central message of
our work.

	If the reader is by any means uneasy
about the existence of a global symmetry, we invite her to
consider gauging it, since it changes none of our
analysis.  Of course, this picks up the anomaly
free     $\tilde U(1)$, say B-L.  The advantage
in this case is that the scale of symmetry breaking
of  B-L   may be kept much lower, all the way to TeV
energy or so.  Furthermore, the B-L version can be
naturally embedded in L-R models or SO (10) grand unified
theory.

	Finally, for laboratory purposes
the situation with  $\tilde U(1)$              explicitly
broken may be even more interesting.  Although, in this
case the objects that carry Z flux are only classically
stable, they may be both long-lived enough and light
enough to be produced in the supercolliders.  Also this is
the minimal Higgs structure that leads to the above
phenomenon and it can result in Z flux tubes being
boundaries of CP domain walls. For this to work, there
must be a pseudoscalar particle with a mass definitely
smaller than M$_W$.
\vskip2truecm

\centerline{\bf Acknowledgments}

The authors would like to thank Professor Abdus Salam, the International
Atomic Energy Agency and UNESCO for hospitality at the International
Centre for Theoretical Physics, Trieste. They are also grateful to Zurab
Berezhani and Balram Rai for useful discussions.
\vfill\eject

\centerline{\bf REFERENCES}
\bigskip\bigskip
\noindent [1]\ \
 H.B. Nielsen and P. Olesen, Nucl. Phys. {\bf B61}, 45 (1973).
\medskip

\noindent [2]\ \
For a review and references, see A. Vilenkin, Phys.
Reports {\bf 121}, 263 (1985).
\medskip

\noindent [3]\ \  T. Vachaspati, Phys. Rev. Lett. {\bf 68}, 1977
(1992).
\medskip

\noindent [4]\ \  M. James, L. Perivolaropoulos and T.
Vachaspati, Phys. Rev. {\bf D46}, R5232\hfill\break
\null\qquad (1992).
\medskip

\noindent [5]\ \  R. Peccei and H., Quinn, Phys. Rev. Lett. {\bf
38}, 1440 (1977).
\medskip

\noindent [6]\ \  C. Hill, A. Kagan and L. Widrow, Phys. Rev.
{\bf D38}, 1100 (1988).
\medskip

\noindent [7]\ \  For a review and references, see J. Kim, Phys.
Reports {\bf 150}, 1 (1987); \hfill\break
\null\qquad H.-Y. Cheng, {\it ibid}. {\bf 158}, 1 (1988).
\medskip

\noindent [8]\ \  A. Zhitnitsky, Sov. J. Nucl. Phys. {\bf 31},
260 (1980);\hfill\break
\null\qquad M. Dine, W. Fischler and M. Srednicki, Phys. lett. {\bf
104B}, 199 (1981).
\medskip

\noindent [9]\ \  A. Vilenkin and A. Everett, Phys. Rev. Lett.
{\bf 48}, 1867 (1982);\hfill\break
\null\qquad T.W. Kibble, G. Lazarides and Q. Shafi, Phys. Rev. {\bf
D26}, 435 (1982).
\medskip

\noindent [10]\ \  J.Y. Chikashige, R.N. Mohapatra and R.D.
Peccei, Phys. Lett. {\bf 98B}, 265 (1981).
\vfill
\end{document}